\documentclass[11pt,a4paper]{article}            
 \usepackage[skins,theorems]{tcolorbox}
\tcbset{highlight math style={enhanced,
  colframe=red,colback=white,arc=0pt,boxrule=1pt}}
  \usepackage[bookmarksopen, bookmarksnumbered, bookmarksopenlevel=2]{hyperref}
  \usepackage{tikz}
  \usepackage{tikz-3dplot}
 \usetikzlibrary{calc}
 \usetikzlibrary{decorations} %
 \usepackage[UKenglish]{babel}
 \usepackage[toc,page]{appendix}
 \usepackage{amsmath}
 \usepackage{amssymb}
 \usepackage{graphicx}
 \usepackage{hhline}
 \usepackage[bf]{caption}
\usepackage{cite}
\usepackage[vcentermath]{youngtab}
\usepackage{geometry}
\usepackage{slashed}
\usepackage{color}
\usepackage{stackrel}
\usepackage{tikz-cd} 
\usepackage{cancel} 
\usepackage{multirow} 
\usepackage{longtable}
\usepackage[all]{xy}

\usepackage{soul, xcolor}

\newtheorem{theorem}{Theorem}%[section]
\newtheorem{lemma}[theorem]{Lemma}

\newtheorem{corollary}[theorem]{Corollary}

\usepackage{scalerel,xcolor}
\def\darrow{\mathrel{\ThisStyle{\ooalign{$\SavedStyle\rightarrow$\cr%
  \hfil\textcolor{white}{\rule{2\LMpt}{1\LMex}}\kern2\LMpt\hfil}}}}

\usepackage{tikz}
\usetikzlibrary{shadows} %defines shadows
\usepackage[framemethod=tikz]{mdframed}

\global\mdfdefinestyle{myboxstyle}{%
  shadow=true,
  linecolor=black,
  shadowcolor=black,
  shadowsize=6pt,
  nobreak=false,
  innertopmargin=10pt,
  innerbottommargin=10pt,
  leftmargin=5pt,
  rightmargin=5pt,
  needspace=1cm,
  skipabove=10pt,
  skipbelow=15pt,
  middlelinewidth=1pt,
  afterlastframe={\vspace{5pt}},
  aftersingleframe={\vspace{5pt}},
  tikzsetting={%
draw=black,
very thick} }

\newmdenv[style=myboxstyle]{whitebox} \newmdenv[style=myboxstyle,backgroundcolor=black!20]{graybox}

\newmdenv[style=myboxstyle,nobreak=true]{blockwhitebox}
\newmdenv[style=myboxstyle,backgroundcolor=black!20,nobreak=true]{blockgraybox}
\newmdenv[nobreak=true,hidealllines=true]{blockbox}

\usepackage{empheq}
\usepackage{arydshln}

\newcommand{\bqa}{\begin{eqnarray}}
\newcommand{\eqa}{\end{eqnarray}}

\newcommand{\nn}{\nonumber}

\renewcommand{\arraystretch}{1.5}

\def\et24{\eta^{24}}
\def\oet24{\frac1{\eta^{24}}}

\hypersetup{
    pdftitle={},
    pdfauthor={},
    pdfsubject={}
}
\numberwithin{equation}{section}
\numberwithin{table}{section}

\makeatletter

\newcommand{\be}{\begin{equation}}
\newcommand{\ee}{\end{equation}}
\newcommand{\beq}{\begin{equation}}
\newcommand{\eeq}{\end{equation}}
\newcommand{\ba}{\begin{aligned}}
\newcommand{\ea}{\end{aligned}}

\newcommand{\bea}{\begin{eqnarray}}
\newcommand{\eea}{\end{eqnarray}}

\newcommand\bi{\begin{itemize}}
\newcommand\ei{\end{itemize}}

\def\unit{{1\kern-.65ex {\rm l}}}
\def\1{{1\kern-.65ex {\rm l}}}

\newcount\hour \newcount\minute
\hour=\time \divide \hour by 60
\minute=\time
\def\now{%
\ifnum \hour<13
  \ifnum \hour=0 \advance \hour by 12 \number\hour:\else \number\hour:\fi%
     \ifnum \minute<10 0\fi%
     \number\minute%
\ A.M.%
\else \advance \hour by -12 \number\hour:%
  \ifnum \minute<10 0\fi%
  \number\minute%
  \ P.M.%
\fi%
}

\makeatother

\begin{document}
\setstcolor{red}

\begin{flushright}
%{\tt\normalsize {CERN-TH-XXX}}\\
%{\tt\normalsize CTPU-PTC-XXX}\\
%{\tt\normalsize ZMP-HH/XXX}
\end{flushright}

\vskip 40 pt
\begin{center}
{\Large \bf 
A Picard rank bound for base surfaces of elliptic Calabi-Yau 3-folds
} 

\vskip 11 mm

{{Caucher Birkar$^{a,}$}\footnote{birkar@tsinghua.edu.cn}}
{and}
{{Seung-Joo Lee$^{b,}$}\footnote{seungjoolee@yonsei.ac.kr}}\\[10pt]

\vspace{0.23cm}
${}^a$ {\it 
Yau Mathematical Sciences Center, Jingzhai, Tsinghua University, \\
Hai Dian District, Bejing, China 100084
}\\[2ex]
${}^b$ {\it 
Department of Physics, Yonsei University, Seoul 03722, Republic of Korea
}\\[2ex]

\end{center}

%\small ${}^{1}${\it Center for Theoretical Physics of the Universe, \\ Institute for Basic Science, Daejeon 34051, South Korea} \\[3 mm]
%\small ${}^{2}${\it CERN, Theory Department, \\ 1 Esplande des Particules, Geneva 23, CH-1211, Switzerland} \\[3 mm]
%\small ${}^{3}${\it II. Institut f\"ur Theoretische Physik, Universit\"at Hamburg, \\  Luruper Chaussee 149, 22607 Hamburg, Germany } \\[3 mm]
%\phantom{\small ${}^{3}$}{\it Zentrum f\"ur Mathematische Physik, Universit\"at Hamburg, \\ Bundesstrasse 55, 20146 Hamburg, Germany  }   \\[3 mm]

%\fnote{}{Email: }

\vskip 5mm

\begin{abstract}\noindent
We compute an explicit rank bound on the Picard group of the compact surfaces, which can serve as the base of an elliptic Calabi-Yau variety with canonical singularities. To bound the Picard rank from above, we develop a novel strategy in birational geometry, motivated in part by the physics of six-dimensional vacua of F-theory as discussed in a companion paper~\cite{steft} to this one. The derivation of the concrete bound illustrates the strategy and clarifies the origin of the boundedness.

\end{abstract}

\vfill

\thispagestyle{empty}
\setcounter{page}{0}
\setcounter{footnote}{0}

\newpage

%\tableofcontents

%\thispagestyle{empty}
%\setcounter{page}{1} 
%\newpage

%%%%%%%%%%
\section{Introduction}\label{sec:intro}
Addressing boundedness is a major goal to achieve in algebraic geometry. A variety of ideas and techniques have been developed to successfully establish boundedness of fibrations~\cite{general}, albeit mostly in an abstract setting. 
Independently in theoretical high energy physics, boundedness of the set of quantum gravity theories was suggested in the context of the swampland program~\cite{Vafa:2005ui} (see also~\cite{Acharya:2006zw, Hamada:2021yxy}). Since a plethora of effective quantum gravity theories arise from geometric compactifications of string theory, compact varieties allowed by string theory, notably, Calabi-Yau varieties, are expected to form a bounded family. 

While boundedness of Calabi-Yau varieties has remained an open issue in complex dimension $d \geq 3$, that of elliptic Calabi-Yau varieties was successfully established in dimension $d=3$~\cite{Gross, Filipazzi}, and more recently, progress has been made in higher dimensions~\cite{45}. 
Elliptic Calabi-Yau varieties play a distinguished role also in string theory. In particular, in F-theory~\cite{Ftheory}, the base of an elliptic fibration is considered an internal space of Type IIB string theory and the fibration econdes the internal profile of the axio-dilation field, thereby fully geometrizing a string vacuum. Accordingly, boundedness of elliptic Calabi-Yau 3-folds amounts to that of the consistent six-dimensional $N=(1,0)$ supergravity theories obtained as F-theory vacua. For applications in physics, it is then important to proceed further to promote the abstract boundedness notion to an explicit one. Notably, explicit upper-bounds on the spectrum of supergravity multiplets, if calculable, would inspire quantum field theorists and model builders alike. 

The aim of this paper is to present an explicit rank bound on the Picard group of the compact surface, which can serve as the base of an elliptic Calabi-Yau 3-fold. Our main theorem is the following.%\footnote{It is worth pointing out that the explicit Picard rank bound of 568 is by no means optimal. After all, as will be manifested along the way, its derivation leaves much room for improvement and the concrete bound obtained should therefore be viewed as a proof of principle. In the physics literature, it was conjectured~\cite{Morrison:2012js} that the maximal Picard rank is $194$, which is realized by a concrete F-theory vacuum~\cite{Aspinwall:1997ye, Candelas:1997eh}. Recently, this stronger bound has been proven physically~\cite{Kim:2024hxe} by imposing the anomaly consistency and some general constraints of quantum gravity. It is thus our belief that we may use the bounding strategy developed in this paper in a more optimal fashion to push the bound down to $194$.}

\begin{theorem}\label{thm0-3}
Assume that 
\bi
\item $(Z,B)$ is a projective $\frac16$-lc pair of dimension 2, 
\item $Z$ is a rational surface,%$Z \to Z_0$ is birational where $Z_0$ is either $\mathbb P^2$ or a Hirzebruch surface $\mathbb F_a$, 
\item $12(K_Z+B)\sim 0$,
\ei
then the Picard number $\rho(Z) \leq 568$. 
\end{theorem}

The bound on the base Picard rank $\rho$ in Theorem~\ref{thm0-3} immediately turns into an explicit bound on the tensor multiplet count, $T$, of the six-dimensional $N=(1,0)$ vacua of F-theory:
\beq
\rho \leq 568 \quad \Leftrightarrow \quad T \leq 567 \,.
\eeq
Moreover, the algebro-geometric strategy we will develop in this paper for the bounding task has a natural counterpart in string theory, and more generally in effective quantum field theory. In this paper, however, we will focus exclusively on the formulations in algebraic geometry. Those readers interested in this connection between algebraic geometry and physics are kindly referred to a companion paper~\cite{steft} to this one for more details. 

The plan of this paper is as follows. In the next section, for the purpose of establishing a Picard rank bound, we introduce a sequence of blowups and graphically visualize it with a tree. Building upon the preliminary results on the blowup sequence, in section~\ref{sec:strategy} we systematically bound the number of blowups involved to prove our main theorem.  \\[-.1in]

\noindent{\it Authors' note:} {\it Having announced part of our results~\cite{FirstTalk} at the stage of completing the work, we learned of the highly relevant paper~\cite{Kim:2024hxe}. It is worth pointing out that the explicit Picard rank bound of 568, as proposed by Theorem~\ref{thm0-3}, is by no means optimal. After all, as will be clear from the proof of the theorem, the derivation of this bound leaves much room for improvement and the concrete number presented should therefore be viewed in part as a proof of principle. It was conjectured~\cite{Morrison:2012js} that the maximal Picard rank is $194$, which had been realized by a concrete F-theory vacuum~\cite{Aspinwall:1997ye, Candelas:1997eh}. In~\cite{Kim:2024hxe} this stronger bound is proven physically via the anomaly constraints, where the authors assume that the classification of superconformal theories and little string theories in~\cite{Heckman:2015bfa, Bhardwaj:2015xxa, Bhardwaj:2015oru, Bhardwaj:2019hhd} is complete and that light strings emerge in the boundary of the tensor branch~\cite{Lee:2019wij}. It is thus our belief that we may use the bounding strategy developed in this paper in a more optimal fashion to push the bound down to $194$.}

\section{Preliminaries}\label{sec:prelim}

\begin{lemma}\label{lem0-1} Assume that 
\bi
\item $X$ is a 3-fold with canonical singularities 
\item $X \xrightarrow[]{h}  Z$ is an elliptic fibration with $K_X \equiv 0$ over $Z$ and with a rational section  
\item $K_X \equiv h^* (K_Z + \Lambda_Z + M_Z)$ is given by the canonical bundle formula where $\Lambda_Z$ and $M_Z$ are the discriminant and the moduli divisors, respectively. 
\ei
Then the pair $(Z, \Lambda_Z + M_Z)$ is $\frac16$-lc. 
\end{lemma}

\paragraph{Proof} In general, a pair $(V, B)$ is $\epsilon$-lc if for any resolution $\varphi: W\to V$, 
\beq
K_W + B_W = \varphi^*(K_V + B) \,,
\eeq
the coefficients of $B_W$ are less than or equal to $1-\epsilon$. It is well-known that to show $(Z, \Lambda_Z+M_Z)$ is $\frac16$-lc, we can take an appropriate resolution of $Z$ and take hyperplane sections to reduce the problem to the case with ${\rm dim}(X)=2$~\cite{birkar}. 
So we can assume from now on that $X \xrightarrow{h} Z$ is an elliptic fibration from a surface onto a curve, with $K_X \equiv 0$ over $Z$ and $h$ equipped with a section. 

Taking a resolution of $X$ we can assume it is smooth. To show that the coefficients of $\Lambda_Z$ at a point $z \in Z$ is less than or equal to $\frac56$ is equivalent to showing that $(X, \frac16 h^*z)$ is lc.\footnote{lc means $0$-lc.} Here, the latter can be checked via the Kodaira classification; see the proof of Lemma~\ref{lem0-2} for more details.  \hfill\(\Box\)

\begin{lemma}\label{lem0-2}
Under the assumption of {\rm Lemma~\ref{lem0-1}}, the coefficients of $\Lambda_Z$ belong to
\beq\label{set}
\{0, \frac16, \frac14, \frac13, \frac12, \frac23, \frac34, \frac56\}\,. 
\eeq \\[-1cm]
\end{lemma}

\paragraph{Proof} As in the proof of Lemma~\ref{lem0-1}, we can reduce to the case ${\rm dim} (X) = 2$. The coefficient of $\Lambda_Z$ at a point $z \in Z$ is equal to $1-t$, where
$t= {\rm sup}\{ s\;|\;(X, s\, h^*z)~\text{is}~{\rm lc} \}$ is the lc-threshold. 
The possible fibers of $h^*z$ are given on page 5 of~\cite{miranda}. We list in Table~\ref{fig:table} the possible fiber types, as well as the corresponding $t$ and the coefficient of $\Lambda_Z$ computed for each fiber type. \hfill\(\Box\)

\begin{table}[t!]
  \centering
  \begin{tabular}{c||c|c}\def\arraystretch{2.2}
                Kodaira type &  lc threshold $t$ & coefficient of $\Lambda_Z$ \\ \hline
                $I_0$ &    1 & 0  \\
                $I_1$ &    1 & 0  \\
                $I_{n\geq 2}$ &  1 & 0 \\ 
                $I_{n\geq 0}^*$ & $\frac12$ & $\frac12$ \\
                $II$ & $\frac56$ & $\frac16$\\ 
                $III$ & $\frac34$ & $\frac14$ \\
                $IV$ & $\frac23$ &$\frac13$ \\
                $IV^*$ & $\frac13$ & $\frac23$\\ 
                $III^*$ & $\frac14$  & $\frac34$\\ 
                $II^*$ & $\frac16$ &$\frac56$ \\ 

\end{tabular}
                \caption{The lc threshold $t$ and the coefficient of $\Lambda_Z$ at $z \in Z$ supporting a Kodaira type fiber
                }
\label{fig:table}
\end{table}

\begin{lemma}\label{lem1/12} 
Assume $X\to Z$ and $(Z, \Lambda_Z + M_Z)$ is as in Lemma~\ref{lem0-1}. Assume also that $K_X \sim 0$. Then, we can find a boundary divisor $B$ such that $12 (K_Z +B)  \sim 0$ and $(Z, B)$ is $\frac16$-lc.

% in the construction~\eqref{chain}, the coefficients in the divisors $B_i$ lie in $\frac{1}{12} \mathbb Z^{\geq 0}$. 
\end{lemma} 

\paragraph{Proof}   
Since $K_X \sim 0$, we also have 
\beq
K_Z + \Lambda_Z + M_Z \sim 0 \,.
\eeq
Let $\bar Z \to Z$ be a high resolution. Then pullback of $K_Z + \Lambda_Z + M_Z$ can be written as 
\beq
K_{\bar Z} + \Lambda_{\bar Z} + M_{\bar Z} \,. 
\eeq
Moreover, $M_{\bar Z} = \frac{1}{12} L_{\bar Z}$ for some Cartier divisor {$L_{\bar Z}$}, which is nef {and free}. 
If we take a general {$0\leq 12 N_{\bar Z} \sim 12 M_{\bar Z}$}, the pair 
$(\bar Z, B_{\bar Z}:=\Lambda_{\bar Z}+N_{\bar Z})$ is sub-$\frac16$-lc {and $12 (K_{\bar Z} + B_{\bar Z}) \sim 0$.}
Let $B$ be the pushdown of $B_{\bar Z}$. 
Then, $12 (K_Z +B) \sim 12 (K_Z + \Lambda_Z + M_Z) \sim 0$ and $(Z, B)$ is $\frac16$-lc.%Then, we can deduce that each coefficient of {\st{$B_Z^{\rm v}$}} {\blue $B_Z$} belongs to $\frac{1}{12}\mathbb Z^{\geq0}$, and upon applying construction~\eqref{chain} {\blue along with~\eqref{blowdown}}, that the same is true of the coefficients of $B_i$. % $h_i = \mu_{z_i} B_i^{\rm h}$ for each $0\leq i <r$. 
\hfill\(\Box\) \\

In view of Lemma~\ref{lem1/12}, it is thus enough to prove Theorem~\ref{thm0-3}. For the purpose of bounding $\rho(Z)$, we can take the minimal resolution and assume that $Z$ is smooth. In the rest of this paper, we consider the blowdown
\beq\label{blowdown}
Z \xrightarrow{\varphi} Z_{0} \,,
\eeq
where $Z_0$ is either $\mathbb P^2$ or a Hirzebruch surface $\mathbb F_a$ for some $a$. But we can ignore the case with $Z_0=\mathbb P^2$ in bounding $\rho(Z)$, since the projective plane $\mathbb P^2$ upon a single blowup already turns into a Hirzebruch surface $\mathbb F_1$. Furthermore, it will shortly become manifest that the case with $a>12$ is not allowed; see Lemma~\ref{lem6}.

Let us now repeatedly blow $Z_0$ up as much as possible,  
\beq\label{chain}
Z_r \xrightarrow{\varphi_{r-1}} Z_{r-1} \to \cdots \to Z_1 \xrightarrow{\varphi_0} Z_0\,,  
\eeq 
with each $\varphi_i$ replacing a point $z_{i} \in Z_{i}$ by an exceptional divisor $E_{i+1} \subseteq Z_{i+1}$. We can then show that $Z_r$ maps to the original base $Z$. This would imply in particular that 
\beq\label{h11Z-bound}
\rho(Z) \leq \rho(Z_r) = 2+ r\,.
\eeq

%Note that each map  appearing in the sequence~\eqref{chain} blows up a point $z_{i-1} \in Z_{i-1}$, resulting in an exceptional divisor $E_{i}\subseteq Z_{i}$. 

The task is therefore to find a uniform upper bound on $r$, for which we will make use of the $\frac16$-lc property of the pair $(Z, B)$. Importantly, this singularity structure is preserved under the birational transformations involved. In other words, each of the divisors $B_i$ of $Z_i$ defined by 
\bea
B_0&=& \varphi_*(B) \,,\\ 
K_{Z_i} + B_i &=& (\varphi_{i-1} \circ \cdots \varphi_0)^*(K_{Z_0}+B_0) \,,
\eea
forms a $\frac16$-lc pair $(Z_i, B_i)$ {with $B_i$ effective}. Here, each $B_i$ splits into the horizontal and the vertical parts as
\beq
B_i = B_i^{\rm h} + B_i^{\rm v} \,,
\eeq
with respect to {the fibration, 
\beq\label{pi-proj}
p_i: Z_i \to \mathbb P^1_{\rm b} \,, 
\eeq
induced from the $\mathbb P^1$-fibration $p_0$ of the Hirzebruch surface $Z_0$ as 
\beq
p_i = p_0 \circ \varphi_0 \cdots \circ \varphi_{i-1} \,.
\eeq
}

While the endpoint $Z_r$ of~\eqref{chain} has a smooth and reduced fiber generically, its special fibers can have more than one components possibly with non-trivial multiplicities. Then, the spirit of our bounding strategy is to constrain the structure of such special fibers, supported over the set of points $x \in X \subset \mathbb P^1_{\rm b}$. To this end, for each special fiber over $x \in X$, we consider the multiplicity 
\beq
\mu_x^{\rm sf} := \mu_x \Lambda_{\mathbb P^1_{\rm b}} \,,  
\eeq
where 
$\Lambda_{\mathbb P^1_{\rm b}}$ is the discriminant divisor appearing in the canonical bundle formula, 
\beq
K_{Z_{r}} + B_{r} \equiv p^*_{r}(K_{\mathbb P^1_{\rm b}} + \Lambda_{\mathbb P^1_{\rm b}} + M_{\mathbb P^1_{\rm b}}) \,, 
\eeq
with the moduli part denoted by $M_{\mathbb P^1_{\rm b}}$. Importantly, the multiplicities $\mu_x^{\rm sf}$ are subject to the global constraint, 
\beq\label{deg-B}
\sum_{x \in X} \mu_x^{\rm sf} = {\rm deg}(\Lambda_{\mathbb P^1_{\rm b}}) \leq 2 \,.
\eeq

For the practical purpose, let us decompose the loci $X$ of special fibers into three disjoint subsets, 
\beq
X = X_1 \cup X_2 \cup X_3 \,,  
\eeq
where 
\bea
X_1 &:=& \{x \in X\;|\; 0\leq \mu_x^{\rm sf} < \frac34 \}\,,\\ 
X_2 &:=& \{x \in X\;|\; \frac34\leq \mu_x^{\rm sf} < \frac56 \}\,,\\ 
X_3 &:=& \{x \in X\;|\; \frac56\leq \mu_x^{\rm sf}  \}\,. 
\eea
Then, the inequality~\eqref{deg-B} constrains the cardinalities of the latter two subsets as
\beq\label{X2X3}
|X_2| + |X_3| \leq 2 \,.   
\eeq

Let us now note that each blowup point $z_i$ lies either in one or two components of the fiber $p_i^{-1}(x)$ with $x:=p_i(z_i)$. Accordingly, we call $\varphi_i$ either a {single}-type or a {double}-type blowup and {designate} the associated exceptional curve as a node of type $\rm I$ and $\rm II$, respectively. Then, the  {exceptional (over $Z_0$)} components in $Z_r$ that map to a given point $z \in Z_0$ via the composition $\varphi_{r-1} \circ \cdots \circ \varphi_0$ can be represented by a tree, say, 

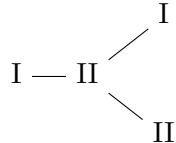
{\centering\begin{tikzpicture}
  \matrix (m) [matrix of nodes,
    row sep = 0.1em, column sep = 1.2em,
    nodes = {text depth = 1ex, text height = 2ex}
  ]
    {
     &  & I \\
      I  & II &  \\
        &   & II \\
  };
  
  \draw [-]
    (m-2-1) edge (m-2-2)
    (m-2-2) edge (m-1-3)
    (m-2-2) edge (m-3-3)
  ;
\end{tikzpicture}\\[-.1in]}
\noindent which means that $z_{i_1}$ is blown up first by a single-type $\varphi_{i_1}$, and that a point $z_{i_2}$ with $i_2 >i_1$ on $E_{i_1+1}$ by a double-type $\varphi_{i_2}$, and that two points on $E_{i_2+1}$, one by a single-type and the other by a double-type blowups; note that our convention is to have a tree grow from left to right via a blowup sequence.

This way, we may visualize the total of $r$ {exceptional} curves of $Z_r$ which map to points of $Z_0$, as disjoint trees sitting on special fibers. 
Obviously, each ``parent'' node of type I can branch at best into one ``child'' node of type II (possibly on top of other child nodes of type I), and a type II parent, into at best two type II children. As for a comment on nomenclature, for two blowups $\varphi_{i_1}$ and $\varphi_{i_2}$ with $i_2>i_1$ generating the nodes $N_{i_1}$ and $N_{i_2}$, respectively, we will say that $N_{i_1}$ is a parent of $N_{i_2}$ and equivalently, that $N_{i_2}$ is a child of $N_{i_1}$, if and only if $z_{i_2} \in E_{i_1+1}$.

For a practical purpose, we define the ``horizontal multiplicity'' $h_i$ of each blowup $\varphi_i$ as  
\beq\label{order-param}
h_i:=\mu_{z_i} B_i^{\rm h}\,, 
\eeq 
which we will also refer to as the horizontal multiplicity of the associated node.  
%\beq\label{subdivide}
%\left\{
%  \begin{array}{ l l}
%    \circ~\, \text{subtype 1}: & \mu_i^{\rm h}\geq \frac16\\ [4pt]
%   \circ~\, \text{subtype 2}: &\mu_i^{\rm h}  =\frac{1}{12} \\[4pt]
%   \circ~\,  \text{subtype 3}: & \mu_i^{\rm h} =0 
%  \end{array}
%\right.
%\eeq
Then, any type I node has its horizontal multiplicity $h \geq \frac16$ due to the $\frac16$-lc property of the pairs $(Z_i, B_i)$ {with $B_i \geq 0$}, while a type II node may as well have $h <\frac16$. 
Note that the only allowed multiplicities in the latter case are $h = \frac{1}{12}$ or $0$.

Without loss of generality, we assume that the series $\{h_i\}_{i=0}^{r-1}$ of horizontal multiplicities is non-increasing, and denote by $r_1$ what distinguishes the two regimes of $h_i$ as follows: 
\beq\label{r'}
h_0 \geq \cdots \geq h_{r_1-1} \geq \frac{1}{6} > h_{r_1} \geq \cdots \geq h_r \,.
\eeq

\section{A Picard Rank Bound}\label{sec:strategy}

Let $\mathcal S_\alpha$ be the total number of type I nodes attached to the special fibers over $X_\alpha$, $\alpha=1,2,3$, and similarly, let $\mathcal D_\alpha$ be that of type II nodes over $X_\alpha$. 

\begin{lemma}\label{lem1} 
The coefficient $b$ of $B_r$ in each component of a special fiber over $x\in X$ is subject to 
\beq\label{b-ineq}
b \leq 
\begin{cases}
    \frac{8}{12} & \text{if } x \in X_1 \\
    \frac{9}{12} & \text{if } x \in X_2  \\  
    \frac{10}{12} & \text{if } x \in X_3 
\end{cases}
\eeq  
\end{lemma} 

\paragraph{Proof} 
For each component $T$ of a special fiber $F$ over $x \in X$, the multiplicity $\mu_x^{\rm sf}$ is subject to the following inequality, 
\beq\label{mu-sf}
\mu_x^{\rm sf} \geq 1-\frac{1-\mu_T B_r}{\mu_T F} \,, 
\eeq
where $b_T :=\mu_T B_r$ is the coefficient of $T$ in $B_r$ and $\mu_T F \in \mathbb Z_{\geq 1}$ is the multiplicity of $T$ in the full fiber. Therefore, if $\mu_x^{\rm sf}$ is strictly bounded from above by $\frac{k}{12}$ with $k\geq \mathbb Z_{\geq 1}$, we have 
\beq
\frac{k}{12} > \mu_x^{\rm sf} \geq b_T ~\Rightarrow~\frac{k-1}{12} \geq b_T \,, \quad \forall T \,, 
\eeq
manifesting the first two inequalities of~\eqref{b-ineq}. On the other hand, the last inequality of~\eqref{b-ineq} is obvious from the $\frac16$-lc property of the pair $(Z_r, B_r)$. \hfill\(\Box\)

\begin{corollary}\label{coro2} 
Each node of type I attached to a special fiber over $x\in X$ has the associated horizontal multiplicity $h$ subject to 
\beq\label{lb}
h \geq 
\begin{cases}
    \frac{4}{12} & \text{if } x \in X_1 \\
    \frac{3}{12} & \text{if } x \in X_2  \\  
    \frac{2}{12} & \text{if } x \in X_3 
\end{cases}
\eeq  
Similarly, $h$ is also subject to the following upper bound, 
\beq\label{ub}
h \leq 
\begin{cases}
    \frac{20}{12} & \text{if } x \in X_1 \\
    \frac{21}{12} & \text{if } x \in X_2  \\  
    \frac{22}{12} & \text{if } x \in X_3 
\end{cases}
\eeq

\end{corollary}

\paragraph{Proof} 
Suppose that a type $I$ node is generated by $\varphi_i$, which blows up a point $z_i$ in a {vertical} curve $C$, leading to an exceptional curve $C':=E_{i+1}$. Then, the coefficients $b:={\rm coeff}_C B_i$ and $b':={\rm coeff}_{C'} B_{i+1}$, are related by 
\beq
b + h -1 = b' \quad \Rightarrow \quad  h= 1 + b' - b \,, 
\eeq
where $h:=h_i$ denotes the horizontal multiplicity~\eqref{order-param} of the single-type blowup. Then, the inequality~\eqref{b-ineq} of Lemma~\ref{lem1} applies both to $b$ and $b'$, from which the bounds~\eqref{lb} and~\eqref{ub} immediately follow. \hfill\(\Box\)

\begin{lemma}\label{lem3} 
For the nodes attached to special fibers over $X_1$, we have 
\beq
\mathcal D_1 \leq 3 \mathcal S_1 \,.
\eeq 
\end{lemma}

\paragraph{Proof}
The inequality~\eqref{mu-sf}, once applied to the multiplicity $\mu_x^{\rm sf}$ of a special fiber $F$ over $x\in X_1$, tells us that $\mu_T F <4$ for every component $T$ of $F$. Therefore, any set of blowups, whose associated nodes form a chain within a given tree of nodes in $F$, can have at best two double-type blowups. Clusters of type II nodes within a tree can therefore come only in the following three forms:\footnote{Hereafter, a cluster within a tree will refer to a connected subtree of the tree.}

{\centering\begin{tikzpicture}
  \matrix (m) [matrix of nodes,
    row sep = 0.1em, column sep = 1.2em,
    nodes = {text depth = 1ex, text height = 2ex}
  ]
    {
  & & &  & II \\
II \,;& II &  II \,; & II &  \\
 &  &     &   & II \\
  };
  
  \draw [-]
    (m-2-2) edge (m-2-3)
    (m-2-4) edge (m-1-5)
    (m-2-4) edge (m-3-5)
  ;
\end{tikzpicture}\\[-.1in]}
\noindent Then, because maximal clusters consisting only of type II nodes, should be branched from a type I parent node, given also that the latter can have at best one type II child node, we learn that $\mathcal D_1 \leq 3 \mathcal S_1$. 
\hfill\(\Box\) \\

For special fibers over $X_{\alpha}$ with $\alpha = 2$ and $3$, it is helpful to refine the numbers $\mathcal S_\alpha$ and $\mathcal D_\alpha$ of type $\rm I$ and $\rm II$ nodes, respectively, by the associated horizontal multiplicities $h$ as follows. Firstly, we denote by $\mathcal S_\alpha^{(12h)}$ the total number of those type I nodes over $X_\alpha$ that are equipped with a fixed multiplicity $h$, where the prefactor $12$ is introduced simply to make the superscript integral. Next, for type II nodes, we refine less and denote collectively by $\mathcal D_\alpha^{(\geq 2)}$ the number of type II nodes with $12 h \geq 2$, which we call type ${\rm II}^{(\geq 2)}$ nodes, and similarly, by $\mathcal D_\alpha^{(< 2)}$, the counterpart for those with $12 h < 2$, which we call type ${\rm II}^{(<2)}$ nodes. 

In what follows we will constrain ${\cal D}_\alpha^{(<2)}$ in terms of ${\cal D}_\alpha^{(\geq 2)}$ and ${\cal S}_\alpha^{(k)}$. To this end,  it proves convenient to consider a one-to-one correspondence, 
\beq\label{Phi}
\Phi_{r_1}:\mathcal W_{r_1} \to \mathcal E_{r_1}\,,
\eeq
between the set, $\mathcal W_{r_1}$, of the $r_1$ points of vertical-curve intersection in $Z_{r_1}$ and the set, $\mathcal E_{r_1}$, of the $r_1$ exceptional curves in $Z_{r_1}$. 
In fact, such a map is naturally defined for each $Z_i$, 
\beq 
\Phi_{i}: \mathcal W_i \to \mathcal E_i \,,\quad  i=1, \cdots, r\,,
\eeq
with $\mathcal W_i$ and $\mathcal E_i$ the counterparts in $Z_i$ of $\mathcal W_{r_1}$ and $\mathcal E_{r_1}$, respectively. The definition is inductive:
\begin{itemize}
\item $\Phi_1$ is trivially defined to map the single point in $\mathcal W_1$ to the exceptional curve $E_1 \in \mathcal E_1$.
\item Given $\Phi_{i}$, to define the next map $\Phi_{i+1}$, we first consider the subset $\mathcal W_{i+1, \,i}\subset \mathcal W_{i+1}$, which collects all points $w\in \mathcal W_{i+1}$ with $\varphi_i(w) \neq z_i$; note that $| \mathcal W_{i+1} \smallsetminus \mathcal W_{i+1,\,i}|=1$ if $\varphi_i$ is a single-type blowup and $2$ if double-type. Each $w \in \mathcal W_{i+1,\,i}$ is then naturally paired with the proper transform of the exceptional curve $\Phi_i(\varphi_i(w))$ to $Z_{i+1}$, while for $w \notin \mathcal W_{i+1,\,i}$, the pairing is defined as follows.  
\begin{enumerate} 
\item If $\varphi_{i}$ is a single-type blowup, the unique point in $\mathcal W_{i+1} \smallsetminus \mathcal W_{i+1,\,i}$ is paired with $E_{i+1}$. 
\item If $\varphi_{i}$ is a double-type blowup, of the two points in $\mathcal W_{i+1} \smallsetminus \mathcal W_{i+1,\,i}$, one  lies in the proper transform of $\Phi_i(z_i)$ to $Z_{i+1}$, so that the two are naturally paired with each other; the other point is then paired with $E_{i+1}$. 
\end{enumerate}
We take the pairing described above between $\mathcal W_{i+1}$ and $\mathcal E_{i+1}$ as the one-to-one correspondence $\Phi_{i+1}$. 
\end{itemize}

Having defined $\Phi_{r_1}$, we now proceed to finding constraints on $D_\alpha^{(<2)}$, for $\alpha=2$ to start with.

% type II node counting is  let $d_2'$ be the number of type II nodes with the associated horizontal multiplicity $h \geq \frac{2}{12}$ and  $d_2 \leq 7 (s_2 + d'_2)$, where . Furthermore, $d'_2$ is subject to $d'_2 \leq 9|X_2|$.

\begin{lemma}\label{lem4} 
For the nodes attached to special fibers over $X_2 \neq \phi$, we have 
\beq\label{D2-ineq-main}
\mathcal D_2^{(<2)} \leq 7 \mathcal D_2^{(\geq 2)} +  \sum_{k=3}^{21}  \sigma_2^{(k)} \mathcal S_2^{(k)} \,, 
\eeq
where $\sigma_2^{(k)}$ are constants subject to 
\beq\label{c2k}
%\begin{cases}
%    \frac{\sigma_2^{(k)}+1}{k}  = \frac{2}{3}  & \text{if } k =12 \\
%    \frac{\sigma_2^{(k)}+1}{k} < \frac{2}{3} & \text{if } k \neq 12  
%    \end{cases}
 {\sigma_2^{(k)}} \leq   k-3 \,\quad ^\forall k= 3, \cdots, 21 \,,
\eeq
and
\beq\label{D2-ineq}
\mathcal D_2^{(\geq 2)} \leq 11 |X_2|  \,.
\eeq

\end{lemma} 

\paragraph{Proof}
Let us consider a maximal cluster $\mathcal C^{(<2)}$ of the nodes with horizontal multiplicities less than $\frac{2}{12}$, sitting within a tree attached to a special fiber over $x \in X_2$. {Then, necessarily, all the nodes in $\mathcal C^{(<2)}$ are of type II$^{(<2)}$, representing the set of blowups over a fixed point $w \in \mathcal W_{r_1} \subset Z_{r_1}$, where two vertical components on $Z_{r_1}$ intersect; we will refer to the exceptional component $E= \Phi_{r_1}(w) \in \mathcal E_{r_1}$, or interchangeably, the corresponding node, as the ``source'' of $\mathcal C^{(<2)}$. }

{Obviously, the source node of $\mathcal C^{(<2)}$ is either of type I or II;}
the horizontal multiplicity $h$ of this {source} node is subject, in the former case, to $\frac{3}{12} \leq h \leq \frac{21}{12}$ by Corollary~\ref{coro2}, and in the latter case, to $h \geq \frac{2}{12}$. %{\st{which follows from the maximality assumption for the cluster $\mathcal C^{(<2)}$.} 
We thus learn that
\beq
\mathcal D_2^{(<2)} \leq \delta_2 \mathcal D_2^{(\geq 2)} +  \sum_{k=3}^{21} \sigma_2^{(k)} \mathcal S_2^{(k)} \,, 
\eeq
for some constants $\delta_2$ and $\sigma_2^{(k)}$; those constants are defined as the maximum number of nodes that such a maximal cluster can ever have, when the source node is of type II$^{(\geq 2)}$ with $h \geq \frac{2}{12}$ and type I with $h=\frac{k}{12}$, respectively.

{In the meantime, the number of nodes in $\mathcal C^{(<2)}$ cannot exceed $7$ in general, which we can confirm as follows. We first note that the cluster $\mathcal C^{(<2)}$ starts from a type II node associated with a double-type blowup, call it $\varphi_{i_0}$ with $i_0 \geq r_1$, blowing up a point $z_{i_0}$ at the intersection of two vertical curves $T$ and $T'$ of $Z_{i_0}$. Note also that one of the two curves here, say $T'$, is the birational transform of the source $E$ of $\mathcal C^{(<2)}$. 
The aforementioned upper-bound of 7 is then realized when the coefficients $b$ and $b'$ in $B_{i_0}$ of the two vertical curves both saturate the coefficient bound~\eqref{b-ineq} and the multiplicity $h_{i_0}$ is also maximal, i.e., when we have
\beq
b=b'=\frac{9}{12}\,,\quad h_{i_0}=\frac{1}{12}\,; 
\eeq
in this case, by performing the maximum number of blowups over $z_{i_0}$, subject to the general constraint that the horizontal multiplicity of a parent node is bigger than or equal to the combined multiplicity of its child nodes, we can indeed construct $\mathcal C^{(<2)}$ with $7$ nodes. 
It thus follows that  
\beq\label{delta-sigma-2-bound}
\delta_2\,, \sigma_2^{(k)} \leq 7 \,, \quad \forall k \,, 
\eeq
confirming the inequality~\eqref{D2-ineq-main} along with the constraint~\eqref{c2k} for $k \geq 10$.}
Then, also for small values of $k$, we can confirm~\eqref{c2k} by calculating $\sigma_2^{(k)}$; see Table~\ref{t:c2k} for the result of this calculation for $k= 3, \cdots, 12$.

\begin{table}
  \centering
  \begin{tabular}{c||c|c||c|c}\def\arraystretch{2.2}
%\begin{table}[h!]
$k:=12 h$      &   max($12b$)     & max($12b'$) & $\sigma_2^{(k)}$ & $k-3$  \\ \hline\hline
~~~~$3$~~~~  & ~~~~~~$9$~~~~~~ & ~~~~~~$0$~~~~~~ & ~~~~$0$~~~~ & ~~~~$ 0$~~~~  \\
$4$  & $9$ & $1$ & $0$ & $ 1$  \\
$5$  & $9$ & $2$ & $1$ & $2$  \\
$6$  & $9$ & $3$ & $1$ & $3$  \\
$7$  & $9$ & $4$ & $2$ & $4$  \\
$8$  & $9$ & $5$ & $2$ & $5$  \\
$9$  & $9$ & $6$ & $3$ & $6$  \\
$10$  & $9$ & $7$ & $4$ & $7$  \\
$11$  & $9$ & $8$ & $5$ & $8$  \\
$12$  & $9$ & $9$ & $7$ & $9$  \\
%$13$  & $8$ & $9$ & $5$ & $10$  \\
%$14$  & $7$ & $9$ & $4$ & $13$  \\
%$15$  & $6$ & $9$ & $3$ & $14$ \\
%$16$  & $5$ & $9$ & $2$ & $15$  \\
%$17$  & $4$ & $9$ & $2$ & $16$  \\
%$18$  & $3$ & $9$ & $1$ & $17$  \\
%$19$  & $2$ & $9$ & $1$ & $18$  \\
%$20$  & $1$ & $9$ & $0$ & $19$  \\
%$21$  & $0$ & $9$ & $0$ & $20$  \\
%\end{table}
\end{tabular}
\caption{ 
The maximum number, $\sigma_2^{(k)}$, of type ${\rm II}^{(<2)}$ nodes constituting a maximal cluster $\mathcal C^{(<2)}$ attached to a special fiber over $X_2$, whose source is a type I node with horizontal multiplicity $h=\frac{k}{12}$. The starting node of the cluster $\mathcal C^{(<2)}$ is associated with the blowup at the intersection of two vertical curves $T$ and $T'$, one of which, say $T'$, is the source of $\mathcal C^{(<2)}$; denoted by $b$ and $b'$ are the coefficients in $B_{r_1}$ of $T$ and $T'$, respectively. }
\label{t:c2k}
\end{table}

{ 
Finally, to derive~\eqref{D2-ineq}, let us first note that each special fiber $F_x$ of $Z_r$ over $x \in X_2$ obeys  
\beq\label{F-D}
B_i^{\rm h} \cdot (\varphi_i \circ \cdots \circ \varphi_{r-1})(F_{x}^{\rm red}) = B_{i+1}^{\rm h} \cdot (\varphi_{i+1} \circ \cdots \circ \varphi_{r-1})(F_{x}^{\rm red}) + \beta_{i,x} \,, 
\eeq
where the superscript ``red'' means that we take the sum of the irreducible components of the fiber $F_x$. Here, the discrepancy $\beta_{i,x}$ coincides with the horizontal multiplicity $h_i$ if $x=p_i(z_i)$ and the $\varphi_i$ is a double-type blowup, and vanishes otherwise. {Note that the relation~\eqref{F-D} works for any $x \in X_1 \cup  X_2 \cup X_3$ and will indeed be used for $x\in X_3$ later on.}

{For the time being, for each $x \in X_2$, it thus follows that }
\beq
 \sum_{i=0}^{r_1-1} \beta_{i,x} < B_{r}^{\rm h} \cdot F_x^{\rm red} + \sum_{i=0}^{r-1} \beta_{i,x} = B_0^{\rm h} \cdot (\varphi_0 \circ \cdots \circ \varphi_{r-1})(F_x^{\rm red}) \leq 2 \,,
\eeq
where we have used  in the first step  that $B_r^{\rm h} \cdot F_x^{\rm red} >0$ and $\beta_{i,x} \geq 0$ in particular for $i \geq r_1$, and in the last step that intersection of $B_0$ with any fiber of $Z_0$ is 2. Denoting by $\mathcal D_{2,x}^{(\geq 2)}$ the number of type II$^{(\geq 2)}$ nodes attached to the special fiber $F_x$, we thus learn that 
\beq
\frac{1}{6} \mathcal D_{2,x}^{(\geq 2)} \leq  \sum_{i=0}^{r_1-1} \beta_{i,x} < 2 \quad \Rightarrow \quad\mathcal D_{2,x}^{(\geq 2)} \leq 11 \,,  
\eeq
as $\beta_{i,x} \geq \frac{1}{6}$ for all $i < r_1$ with $\beta_{i,x} \neq 0$, and hence, that 
\beq
\mathcal D_{2}^{(\geq 2)} = \sum_{x\in X_2} \mathcal D_{2,x}^{(\geq 2)} \leq 11|X_2| \,, 
\eeq
establishing~\eqref{D2-ineq}. }
\hfill\(\Box\) \\

Turning our attention to the special fibers over $X_3$, we will now constrain $\mathcal D_3^{(<2)}$ in a similar fashion to Lemma~\ref{lem4}. For a special fiber $F_x$ over $x\in X_3$, however, we {have not derived} a definite upper bound on $\mu_x^{\rm sf}$, and hence, some of the constraints which used to be available no longer holds; see e.g. Lemma~\ref{lem1}.
In order to more efficiently constrain $\mathcal D_3^{(<2)}$ than how we naively did for $\mathcal D_2^{(<2)}$, we will therefore make {the following refinement of $\mathcal D_3^{(\geq 2)}$, }

\beq\label{def-D-refine}
\mathcal D_3^{(\geq 2)} = {\check{\mathcal D}_3^{(\geq 2)}} + {\bar{\mathcal D}}_3^{(\geq 2)}+ {\hat{\mathcal D}}_3^{(\geq 2)} \,,  
\eeq
based on the properties of the maximal clusters $\mathcal C_{i_m}^{(<2)}$ of type II$^{(<2)}$ nodes, $m=1, \cdots, \mathcal D_3^{(\geq 2)}$, each generated by blowups over the point $w_{i_m+1} \in \mathcal W_{r_1}$ of $Z_{r_1}$ with a double-type blowup $\varphi_{i_m+1}$, as matched by~\eqref{Phi}. Here, the checked, barred, and hatted variables on the right hand side count those clusters $\mathcal C_{i_m}^{(<2)}$ containing $0$, $1$, and more than $1$ nodes with $h=\frac{1}{12}$, respectively. 
Similarly, we also view $\mathcal S_3^{(k)}$ as counting the maximal clusters $\mathcal C_{j_n}^{(<2)}$ of type II$^{(<2)}$ nodes, $n=1, \cdots, \mathcal S_3^{(k)}$, generated over the point $w_{j_n+1} \in \mathcal W_{r_1}$ of $Z_{r_1}$ with a single-type blowup $\varphi_{j_n+1}$, to make the analogous refinement, 
\beq\label{def-S-refine}
 \mathcal S_3^{(k)} = {\check{\mathcal S}}_3^{(k)} +  {\bar{\mathcal S}}_3^{(k)}+ {\hat{\mathcal S}}_3^{(k)} \,. 
\eeq
{As before, the checked, barred, and hatted variables count those $\mathcal C_{j_n}^{(<2)}$ containing 0, 1, and more than 1 nodes with $h=\frac{1}{12}$, respectively.}

For a refined analogue of Lemma~\ref{lem4}, we then constrain $\mathcal D_3^{(<2)}$ as follows. 

\begin{lemma}\label{lem5} 
For the special fibers over $X_3 \neq \phi$, %let us assume that 
%\beq\label{D31-vanish}
%\mathcal D_3^{(1)}=0 \,,
%\eeq
%i.e., that no type II$^{(<2)}$ nodes have horizontal multiplicity $h=\frac{1}{12}$. Then 
we have 
\beq\label{D3-ineq-main}
\mathcal D_3^{(<2)} \leq 11 {\check{\mathcal D}}_3^{(\geq 2)} + 15  {\bar{\mathcal D}}_3^{(\geq 2)} +  22 {\hat{\mathcal D}}_3^{(\geq 2)} + \sum_{k=2}^{22} \big(\check{\sigma}_3^{(k)} \check{\mathcal S}_3^{(k)} + { \bar{\sigma}_3^{(k)}{\bar{\mathcal S}_3}^{(k)} + \hat{\sigma}_3^{(k)}{\hat{\mathcal S}_3}^{(k)}}\big) \,, 
\eeq
where $\check{\sigma}_3^{(k)}$, {$\bar{\sigma}_3^{(k)}$, and $\hat{\sigma}_3^{(k)}$} are constants subject, respectively, to 

\bea\label{c3k-check}
%\begin{cases}
%    \frac{\check{\sigma}_3^{(k)}+1}{k}  = 1 & \text{if } k =12 \\
%    \frac{\check{\sigma}_3^{(k)}+1}{k}  < 1 & \text{if } k \neq 12 
%    \end{cases}
\check{\sigma}_3^{(k)} &\leq& k-1 \,,\\ \label{c3k-bar} 
\bar{\sigma}_3^{(k)} &\leq& k+3 \,, \\ \label{c3k-hat}
\hat{\sigma}_3^{(k)} &\leq& k+10 \,,
\eea
for $k=2, \cdots, 22$.
Furthermore, we have 
\beq\label{D3-ineq}
\check{\mathcal D}_3^{(\geq 2)} + \frac{3}{2} \bar{\mathcal D}_3^{(\geq 2)} + 2 \hat{\mathcal D}_3^{(\geq 2)} + \sum_{k=2}^{22}{ \big(\frac{1}{2} {\bar{\mathcal S}}_3^{(k)} +  {\hat{\mathcal S}}_3^{(k)}\big)} \leq  {\frac{23}{2}} |X_3|  \,.
\eeq

\end{lemma}

\paragraph{Proof}
The proof of this lemma goes in an analogous manner to that of Lemma~\ref{lem4}; we will thus be briefer here. 

Let us consider maximal clusters of type II$^{(<2)}$ nodes over $x\in X_3$. {Each such maximal cluster $\mathcal C^{(<2)}$ must be generated over a point $w \in \mathcal W_{r_1} \subset Z_{r_1}$, where two vertical components of $Z_{r_1}$ intersect. Then, the ``source'' node of $\mathcal C^{(>2)}$, defined as before via the correspondence $E= \Phi_{r_1}(w) \in \mathcal E_{r_1}$, is either of type I with $\frac{2}{12} \leq h \leq \frac{22}{12}$ (see Coroallary~\ref{coro2}) or of type II$^{(\geq 2)}$. }
We thus learn that
\beq
\mathcal D_3^{(<2)} \leq \check\delta_3 \check{\mathcal D}_3^{(\geq 2)} +\bar\delta_3 \bar{\mathcal D}_3^{(\geq 2)}+\hat\delta_3 \hat{\mathcal D}_3^{(\geq 2)}+  \sum_{k=2}^{22} \big(\check{\sigma}_3^{(k)} \check{\mathcal S}_3^{(k)} + {\bar{\sigma}_3^{(k)}{{{\bar{\mathcal S}}}_3}^{(k)}+\hat{\sigma}_3^{(k)}{{{\hat{\mathcal S}}}_3}^{(k)}} \big)\,. 
\eeq
where the various constants $\delta$ and $\sigma$ are defined as follows. Firstly, $\check{\sigma}_3^{(k)}$, {$\bar{\sigma}_3^{(k)}$, and $\hat{\sigma}_3^{(k)}$} denote the maximum number of nodes that such a maximal cluster can ever have, when the source node is of type I with $h=\frac{k}{12}$, under the following assumptions, respectively, that no nodes, precisely one, and more than one nodes of the cluster have $h=\frac{1}{12}$. Similarly, their counterpart for the cases with a type II$^{(\geq 2)}$ {source} node also comes in the refined versions, $\check \delta_3$, $\bar \delta_3$ and $\hat \delta_3$, which respectively, amount to the maximal node count under the assumption that no nodes, precisely one, and more than one nodes in the cluster have $h=\frac{1}{12}$.  

Then, with a type II source node, we have
\beq\label{delta-check-3}
\check \delta_3 \leq 11 \,. 
\eeq
This is realized when the cluster starts from a type II node associated with a double-type $\varphi_{i_0}$ with $i_0 \geq r_1$, blowing up a point $z_{i_0}$ at the intersection of two vertical curves $T$ and $T'$ with coefficients $b$ and $b'$ in $B_{i_0}$ given as 
\beq\label{bb'56}
b=b'=\frac56\,. 
\eeq
Similarly, we can deduce that 
\beq\label{15-22}
\bar \delta_3 \leq 15 \,, \quad\hat \delta_3 \leq 22\,,
\eeq
where the inequalities are saturated once again for~\eqref{bb'56}. 

Next, with a type I source node, {we can calculate $\sigma_3^{(k)}$ in their three versions under the respective refining assumptions, as a refined analogue of $\sigma_2^{(k)}$ in Table~\ref{t:c2k}. %Furthermore, having defined $\check{\sigma}_3^{(k)}$ as the maximal node count under the extra assumption that no nodes in the cluster have $h=\frac{1}{12}$, we can also calculate them easily. 
{See Table~\ref{t:c3k} for the result of this calculation {for $k=2, \cdots, 12$}, which confirms that the constraints~\eqref{c3k-check}, \eqref{c3k-bar} and~\eqref{c3k-hat} are obeyed. Note that the upper bounds of $11$, $15$ and $22$, as presented in~\eqref{delta-check-3} and~\eqref{15-22} for $\check \delta_3$, $\bar \delta_3$, and $\hat \delta_3$, also apply, respectively, to $\check{\sigma}_3^{(k)}$, $\bar{\sigma}_3^{(k)}$, and $\hat{\sigma}_3^{(k)}$ for every $k$; this implies that the three constraints are automatically satisfied for $k>12$. 

}

\begin{table}
  \centering
  \begin{tabular}{c||c|c||c|c||c|c||c|c}\def\arraystretch{2}
%\begin{table}[h!]
$k:=12 h$      &   max($12b$)     & max($12b'$) & $\check\sigma_3^{(k)}$ & $k-1$ &$\bar{\sigma}_3^{(k)}$ &$k+3$& $ \hat{\sigma}_3^{(k)}$ & $k+10$  \\ \hline\hline
~~~~$2$~~~~  & ~~~~~~$10$~~~~~~ & ~~~~~~$0$~~~~~~ & ~~~~$0$~~~~ & ~~~~$1$~~~~ & ~~~~0~~~~ & ~~~~5~~~~ & ~~~~$0$~~~~ & ~~~~$ 12$~~~~  \\ 
$3$  & $10$ & $1$ & $0$ & $2$ &1&6& $1$ & $13$ \\
$4$  & $10$ & $2$ & $1$ & $3$ &1&7& $2$ & $14$  \\
$5$  & $10$ & $3$ & $1$ & $4$ &2&8& $3$ & $15$   \\
$6$  & $10$ & $4$ & $2$ & $5$ &2&9& $4$ & $16$ \\
$7$  & $10$ & $5$ & $2$ & $6$ &3&10& $5$ & $17$ \\
$8$  & $10$ & $6$ & $3$ & $7$ &3&11& $6$ & $18$ \\
$9$  & $10$ & $7$ & $4$ & $8$ &5&12& $8$ & $19$ \\
$10$  & $10$ & $8$ & $5$ & $9$ &6&13& $10$ & $20$ \\
$11$  & $10$ & $9$ & $7$ & $10$ &9&14& $14$ & $21$ \\
$12$  & $10$ & $10$ & $11$ & $11$ &15&15& $22$ & $22$ \\
%$13$  & $9$ & $10$ & $7$ & $12$ &9&18& $14$ & $24$ \\
%$14$  & $8$ & $10$ & $5$ & $13$ &6&19& $10$ & $25$ \\
%$15$  & $7$ & $10$ & $4$ & $14$ &5&20& $8$ & $26$\\
%$16$  & $6$ & $10$ & $3$ & $15$  &3&21& $6$ & $27$\\
%$17$  & $5$ & $10$ & $2$ & $16$  &3&22& $5$ & $28$\\
%$18$  & $4$ & $10$ & $2$ & $17$  &2&23& $4$ & $29$ \\
%$19$  & $3$ & $10$ & $1$ & $18$  &2&24& $3$ & $30$\\
%$20$  & $2$ & $10$ & $1$ & $19$ &1&25& $2$ & $31$ \\
%$21$  & $1$ & $10$ & $0$ & $20$ &1&26& $1$ & $32$ \\
%$22$  & $0$ & $10$ & $0$ & $21$ &0&27& $0$ & $33$ \\
%\end{table}
\end{tabular}
\caption{ 
The maximum number, $\sigma_3^{(k)}$, of type ${\rm II}^{(<2)}$ nodes constituting a maximal cluster $\mathcal C^{(<2)}$ attached to a special fiber over $X_3$, whose source is a type I node with horizontal multiplicity $h=\frac{k}{12}$; the maximum $\sigma_3^{(k)}$ for each $k$ comes in three different versions, $\check\sigma_3^{(k)}$, $\bar{\sigma}_3^{(k)}$, and $\hat{\sigma}_3^{(k)}$, respectively, for which $\mathcal C^{(<2)}$ is assumed to contain 0, 1, and more than 1 nodes with $h=\frac{1}{12}$.
The starting node of the cluster $\mathcal C^{(<2)}$ is associated with the blowup at the intersection of two vertical curves $T$ and $T'$, one of which, say $T'$, is the source of $\mathcal C^{(<2)}$; denoted by $b$ and $b'$ are the coefficients in $B_{r_1}$ of $T$ and $T'$, respectively. }
\label{t:c3k}
\end{table}

\begin{comment}
\begin{table}
  \centering
  \begin{tabular}{c||c|c|c|c}\def\arraystretch{2.2}
%\begin{table}[h!]
$k:=12 h$      &   max($12b$)     & max($12b'$) & $ \sigma_3^{(k)}$ & $\frac{\sigma_3^{(k)}+1}{k}$  \\ \hline\hline
~~~~~~$2$~~~~~~  & ~~~~~~$10$~~~~~~ & ~~~~~~$0$~~~~~~ & ~~~~~~$0$~~~~~~ & ~~~~~~$0$~~~~~~  \\
$3$  & $10$ & $1$  &$1$ & $\frac{2}{3}$ \\
$4$  & $10$ & $2$   &$2$ & $\frac{3}{4}$  \\
$5$  & $10$ & $3$  &$3$ & $\frac{4}{5}$   \\
$6$  & $10$ & $4$  &$4$ & $\frac{5}{6}$ \\
$7$  & $10$ & $5$   &$5$ & $\frac{6}{7}$ \\
$8$  & $10$ & $6$   &$6$ & $\frac{7}{8}$ \\
$9$  & $10$ & $7$   &$8$ & $1$ \\
$10$  & $10$ & $8$  &$10$ & $\frac{11}{10}$ \\
$11$  & $10$ & $9$   &$14$ & $\frac{15}{11}$ \\
$12$  & $10$ & $10$   &$22$ & $\frac{23}{12}$ \\
$13$  & $9$ & $10$   &$14$ & $\frac{15}{13}$ \\
$14$  & $8$ & $10$   &$10$ & $\frac{11}{14}$ \\
$15$  & $7$ & $10$   &$8$ & $\frac{3}{5}$\\
$16$  & $6$ & $10$   &$6$ & $\frac{7}{16}$\\
$17$  & $5$ & $10$    &$5$ & $\frac{6}{17}$\\
$18$  & $4$ & $10$  &$4$ & $\frac{5}{18}$ \\
$19$  & $3$ & $10$  &$3$ & $\frac{4}{19}$\\
$20$  & $2$ & $10$   &$2$ & $\frac{3}{20}$ \\
$21$  & $1$ & $10$  &$1$ & $\frac{2}{21}$ \\
$22$  & $0$ & $10$  &$0$ & $0$ \\
%\end{table}
\end{tabular}
\caption{ \label{t:c3k}}
\end{table}
\end{comment}

{
Finally, the derivation of~\eqref{D3-ineq} also goes similarly to that of~\eqref{D2-ineq} in Lemma~\ref{lem4}. In much the same way as how we derived~\eqref{F-D}, we first note that each special fiber $F_x$ of $Z_r$ over $x \in X_3$ obeys,    
\beq\label{F-D3}
B_i^{\rm h} \cdot (\varphi_i \circ \cdots \circ \varphi_{r-1})(F_{x}^{\rm red}) = B_{i+1}^{\rm h} \cdot (\varphi_{i+1} \circ \cdots \circ \varphi_{r-1})(F_{x}^{\rm red}) + \beta_{i,x} \,, 
\eeq
where the superscript ``red'' again indicates the sum of the irreducible components of the fiber $F_x$. Furthermore, as before, non-trivial values of $\beta_{i,x}$ can only arise as $\beta_{i,x} = h_i$ if $x=p_i(z_i)$ with $\varphi_i$ of a double type. Once again, it thus follows for each $x \in X_3$ that 
\beq\label{D2-source}
 \sum_{i=0}^{r_1-1} \beta_{i,x} < B_{r}^{\rm h} \cdot F_x^{\rm red} + \sum_{i=0}^{r-1} \beta_{i,x} = B_0^{\rm h} \cdot (\varphi_0 \circ \cdots \circ \varphi_{r-1})(F_x^{\rm red}) \leq 2 \,,
\eeq
and hence, that 
\beq\label{D2-11}
\frac{1}{6} \mathcal D_{3,x}^{(\geq 2)} \leq  \sum_{i=0}^{r_1-1} \beta_{i,x} < 2 \quad \Rightarrow \quad\mathcal D_{3,x}^{(\geq 2)} \leq 11 \,.   
\eeq
However, we can easily improve~\eqref{D2-11} in terms of the refined counts~\eqref{def-D-refine} and~\eqref{def-S-refine}. To this end, let us simply note that 
\beq\label{extra}
\frac{1}{12}\bar{\mathcal D}_{3,x}^{(\geq 2)} + \frac{1}{6} \hat{\mathcal D}_{3,x}^{(\geq 2)} + \sum_{k=2}^{22}\big({ {\frac{1}{12}\bar{\mathcal S}_{3,x}^{(k)}} +\frac{1}{6} {\hat{\mathcal S}_{3,x}^{(k)}}}\big)  \leq \sum_{i=r_1}^{r-1} \beta_{i,x} \,, 
\eeq
with the subscripted $x$ indicating that the variable counts the nodes locally for the fixed fiber over $x \in X_3$. Here, we have used that the maximal clusters $\mathcal C_{i_m}^{(<2)}$ defined after~\eqref{def-D-refine} have 1 and more than 1 nodes with $h=\frac{1}{12}$, respectively, if counted by the barred and the hatted variables. It thus follows that 
\bea\nn
\frac16 \big(\check{\mathcal D}_{3,x}^{(\geq 2)} + \frac{3}{2} \bar{\mathcal D}_{3,x}^{(\geq 2)} + 2 \hat{\mathcal D}_{3,x}^{(\geq 2)} &+& \sum_{k=2}^{22}\big({ {\frac{1}{2}\bar{\mathcal S}_{3,x}^{(k)}} + {\hat{\mathcal S}_{3,x}^{(k)}}}\big) \big)  \\ 
&\leq& \frac{1}{6} \mathcal D_{3,x}^{(\geq 2)} +  \big(\frac{1}{12}\bar{\mathcal D}_{3,x}^{(\geq 2)} + \frac{1}{6} \hat{\mathcal D}_{3,x}^{(\geq 2)} +  \sum_{k=2}^{22}\big({ {\frac{1}{12}\bar{\mathcal S}_{3,x}^{(k)}} +\frac{1}{6} {\hat{\mathcal S}_{3,x}^{(k)}}}\big) \big)  \\ 
&\leq& \sum_{i=0}^{r_1-1} \beta_{i,x} +  \sum_{i=r_1}^{r-1} \beta_{i,x} \\ 
 &<& B_{r}^{\rm h} \cdot F_x^{\rm red} + \sum_{i=0}^{r-1} \beta_{i,x} \leq 2 \,,
\eea
where, we have used the refinement~\eqref{def-D-refine} in the first line, the left-most inequality in~\eqref{D2-11} as well as the observation~\eqref{extra} in the second, and~\eqref{D2-source} in the third. We thus learn that 
\beq
\check{\mathcal D}_{3,x}^{(\geq 2)} + \frac{3}{2} \bar{\mathcal D}_{3,x}^{(\geq 2)} + 2 \hat{\mathcal D}_{3,x}^{(\geq 2)} + \sum_{k=2}^{22} \big({ {\frac{1}{2}\bar{\mathcal S}_{3,x}^{(k)}} +{\hat{\mathcal S}_{3,x}^{(k)}}}\big) \leq  { \frac{23}{2}} \,,
\eeq
which, once summed over $x \in X_3$, establishes~\eqref{D3-ineq}.
}
\hfill\(\Box\) \\

\begin{lemma}\label{lem6}
Assume $(Z,B)$ is an $\epsilon$-lc pair of dimension 2 and that $Z$ is smooth projective with $K_Z+B\equiv 0$. Then for any smooth projective curve $C \subseteq Z$, we have
\beq
C^2 \geq -\frac{2}{\epsilon} \,. 
\eeq
In particular, for the case of $\frac{1}{6}$-lc pair $(Z_r, B_r)$, we have $C^2 \geq -12$ for any smooth projective curve $C \subseteq Z_r$. 
\end{lemma}

\paragraph{Proof} Let $b$ be the coefficient of $C$ in $B$. We can assume $C^2 <0$. Then
\beq
(K_Z + B+(1-b) C) \cdot C \geq (K_Z + bC + (1-b)C )\cdot C \,,
\eeq
from which it follows that 
\beq
(1-b) C^2 \geq 2g(C) - 2 \geq -2 \,.
\eeq
Since $(Z,B)$ is $\epsilon$-lc and $C^2<0$, we see that  
\beq
\epsilon C^2 \geq (1-b) C^2 \geq -2 \,,
\eeq
and hence that  
\beq
C^2 \geq -\frac{2}{\epsilon} \,.
\eeq 
In particular, if $\epsilon = \frac{1}{6}$, then we have 
\beq
C^2 \geq -12 \,,
\eeq
for any smooth projective curve $C$. 
\hfill\(\Box\)

\begin{lemma}\label{lem7}
\bea
-K_{Z_0} \cdot B_0^{\rm h} &\leq& 8 \,,\\
-K_{Z_r} \cdot B_r^{\rm h} &\geq& -20 \,.  
\eea
\end{lemma}

\paragraph{Proof} 
For $Z_0 = \mathbb F_a$ a Hirzebruch surface, 
\beq\label{KZ^2}
-K_{Z_0} \cdot B_0^{\rm h} = B_0 \cdot B_0^{\rm h} \leq B_0 \cdot (B_0^{\rm h} + B_0^{\rm v})= B_0^2  = K_{Z_0}^2 = 8 \,, 
\eeq
where, in the inequality, we have used that the vertical part $B_0^{\rm v}$ is a nef divisor. Furthermore, the last equality can be seen from the fact that $K_{Z_0} \sim -(a+2) f - 2e$, where $f$ is a general fiber of $p_0: Z_0 \to \mathbb P^1_{\rm b}$ and $e$ the section with $e^2=-a$. 

Next, to show that 
\beq
-K_{Z_r} \cdot B_r^{\rm h} \geq -20 \,,  
\eeq
let us first recall that each coefficient of $B_r$ is less than or equal to $\frac56$ since $(Z_r, B_r)$ is $\frac16$-lc. For any curve $C \subseteq Z_r$, it then follows that $B_r^{\rm h} \cdot C \geq -10$. To see this we may assume $C^2 <0$ and observe that 
\beq
2p_a(C) - 2 = (K_{Z_r} + C) \cdot C \leq (K_{Z_r} +B_r +(1-b)C) \cdot C < 0 \,, 
\eeq
where $b$ denotes the coefficient of $C$ in $B_r$ and $p_a(C)$ the arithmetic genus of $C$. Hence, $C\simeq \mathbb P^1$. In particular, $C$ is smooth, so $C^2 \geq -12$ by Lemma~\ref{lem6}. 
On the other hand, 
\beq
B_r^{\rm h} \cdot C \geq aC^2 \geq \frac56 C^2 \geq \frac56 \times (-12) =-10 \,,
\eeq
where $a$ is the coefficient of $C$ in $B_r^{\rm h}$. 
Thus, 
\beq
B_r^{\rm h} + 10F 
\eeq
is a nef divisor where $F$ is a general fiber of $p_r:Z_r \to \mathbb P^1_{\rm b}$. 
We therefore have 
\bea
-K_{Z_r} \cdot B_r^{\rm h} &=& -K_{Z_r} \cdot (B_r^{\rm h} + 10F) + K_{Z_r} \cdot 10F \\
&=& B_r \cdot (B_r^{\rm h} + 10F) + K_{Z_r} \cdot 10 F \\ 
&\geq& K_{Z_r} \cdot 10F = -20 \,,  
\eea 
and hence the lemma holds. \hfill\(\Box\)

\begin{lemma}\label{lem8}
\beq\label{336}
4 \mathcal S_1 + \Big(2 \mathcal D_2^{(\geq 2)}  + \sum_{k=3}^{21} k\, \mathcal S_2^{(k)} \Big) +  \Big(2 \mathcal D_3^{(\geq 2)} +  \sum_{k=2}^{22} k\, \mathcal S_3^{(k)} \Big)  \leq 336\,.
\eeq

In fact, 
\beq\label{336-improved}
4 \mathcal S_1 + \Big(2 \mathcal D_2^{(\geq 2)}  + \sum_{k=3}^{21} k\, \mathcal S_2^{(k)} \Big) +  \Big(2 \check{\mathcal D}_3^{(\geq 2)} + 3 \bar{\mathcal D}_3^{(\geq 2)} + 4 \hat{\mathcal D}_3^{(\geq 2)} +  \sum_{k=2}^{22} \big(k \,\check{\mathcal S}_3^{(k)} +  { (k+1) \,{{\bar{\mathcal S}}_3}^{(k)} + (k+2) \,{{\hat{\mathcal S}}_3}^{(k)}} \big)\Big)  \leq 336\,, 
\eeq
where the refined variables defined in~\eqref{def-D-refine} and~\eqref{def-S-refine} have been used.

\end{lemma}

\paragraph{Proof} 
Let us first observe that for each $i = 0, \cdots, r-1$, we have 
\bea
- K_{Z_{i+1}} \cdot B_{i+1}^{\rm h} &=& - (\varphi_{i}^* K_{Z_{i}} + E_{i+1}) \cdot B_{i+1}^{\rm h} \\ 
&=& -\varphi_{i}^* K_{Z_{i}} \cdot B_{i+1}^{\rm h} - E_{i+1} \cdot B_{i+1}^{\rm h} \\ 
&=&  -K_{Z_{i}} \cdot B_{i}^{\rm h} - \mu_{z_i} B_i^{\rm h} \\ \label{inductive}
&=& - K_{Z_{i}} \cdot B_{i}^{\rm h} - h_i \,. 
\eea
where in the last step the definition~\eqref{order-param} for the horizontal multiplicity has been used. 
Then, Lemma~\ref{lem7}, once combined with the relations~\eqref{inductive}, leads to 
\beq
8 \geq - K_{Z_0} \cdot B_0^{\rm h} =  - K_{Z_r}\cdot B_r^{\rm h} + \sum_{i=0}^{r-1} h_i  \geq -20 + \sum_{i=0}^{r-1} h_i \quad \Rightarrow\quad  \sum_{i=0}^{r-1} h_i \leq 28 \,. 
\eeq
Recalling from Corollary~\ref{coro2} that the horizontal multiplicity $h$ of a single-type blowup in $p_r^{-1}(x)$ for $x\in X_1$ is subject to $h \geq \frac{4}{12}$, it follows immediately that 
\beq\label{before-strengthen}
\frac{4}{12} \mathcal S_1 + \Big(\frac{2}{12} \mathcal D_2^{(\geq 2)} + \sum_{k=3}^{21}  \frac{k}{12} \mathcal S_2^{(k)} \Big)  + \Big(\frac{2}{12} \mathcal D_3^{(\geq 2)} + \sum_{i=2}^{22} \frac{k}{12} \mathcal S_3^{(k)}\Big) \leq \sum_{i=0}^{r-1} h_i \leq 28 \,, 
\eeq
ane hence~\eqref{336} holds. 

Note that we may strengthen the constraint~\eqref{before-strengthen} by taking also into account the blowups with $h=\frac{1}{12}$. In particular, for the special fibers over $X_3$, we may further consider the $\bar{\mathcal D}_3^{(\geq 2)}+2 \hat{\mathcal D}_3^{(\geq 2)} { + \sum_{k=2}^{22}  \big(\bar{\mathcal S}_3^{(k)}+2 \hat{\mathcal S}_3^{(k)}\big)} $ blowups with $h=\frac{1}{12}$. This allows us to replace $\frac{2}{12} \mathcal D_3^{(\geq 2)}$ and $\frac{k}{12}\mathcal S_3^{(k)}$ in the left hand side of~\eqref{before-strengthen}, respectively, by 
\beq
\frac{2}{12} \check{\mathcal D}_3^{(\geq 2)} + \frac{3}{12}\bar{\mathcal D}_3^{(\geq 2)} + \frac{4}{12} \hat{\mathcal D}_3^{(\geq 2)}\,\quad\text{and}\quad 
\frac{k}{12} \check{\mathcal S}_3^{(\geq 2)} + \frac{k+1}{12}\bar{\mathcal S}_3^{(\geq 2)} + \frac{k+2}{12} \hat{\mathcal S}_3^{(\geq 2)}\,,
\eeq
establishing the improved constraint~\eqref{336-improved}. 
\hfill\(\Box\)

\paragraph{Proof of Theorem~\ref{thm0-3}} 
In order to establish $\rho(Z) \leq 568$, we will prove that $r \leq 566$ in the construction~\eqref{chain}.   
From Lemmas~\ref{lem3},~\ref{lem4} and~\ref{lem5}, let us first observe that 
\bea
r &= & (\mathcal D_1 + \mathcal S_1) +  (\mathcal D_2^{(\geq 2)} + \mathcal D_2^{(<2)} + \mathcal S_2) + (\mathcal D_3^{(\geq 2)} + \mathcal D_3^{(<2)} + \mathcal S_3)  \\ \nn
&\leq & 4 \mathcal S_1 + 8 \mathcal D_2^{(\geq 2)} + \sum_{k=3}^{21} (\sigma_2^{(k)}+1) \mathcal S_2^{(k)} +12  {\check{\mathcal D}}_3^{(\geq 2)} + 16 {\bar{\mathcal D}}_3^{(\geq 2)} + 23 {\hat{\mathcal D}}_3^{(\geq 2)}   \\ 
&& +\; \sum_{k=2}^{22} \big((\check\sigma_3^{(k)}+1)  \check{\mathcal S}_3^{(k)} { + (\bar\sigma_3^{(k)}+1)  \bar{\mathcal S}_3^{(k)} +(\hat\sigma_3^{(k)}+1)  \hat{\mathcal S}_3^{(k)}}  \big) \\ \nn 
&\leq &  4 \mathcal S_1 + 8 \mathcal D_2^{(\geq 2)} + \sum_{k=3}^{21} { (k-2)} \mathcal S_2^{(k)} +12  {\check{\mathcal D}}_3^{(\geq 2)} + 16 {\bar{\mathcal D}}_3^{(\geq 2)} + 23 {\hat{\mathcal D}}_3^{(\geq 2)}  \\ \label{rmax}
&&+\; \sum_{k=2}^{22} \big(k \check{\mathcal S}_3^{(k)}  + { (k+4)}\bar{\mathcal S}_3^{(k)} + { (k+11)} \hat{\mathcal S}_3^{(k)} \big) \\
&=:& r_{\rm bound} \,,
\eea
where we have used, in the last step before defining $r_{\rm bound}$, that the constants $\sigma_2^{(k)}$, $\check\sigma_3^{(k)}$, $\bar\sigma_3^{(k)}$ and $\hat\sigma_3^{(k)}$ are subject to the inequalities~\eqref{c2k},~\eqref{c3k-check}, ~\eqref{c3k-bar} and~\eqref{c3k-hat}, respectively. 
On the other hand, upon combining the two inequalities~\eqref{D2-ineq} and~\eqref{D3-ineq} with weight 6 and 10, we obtain 
\beq
6 \mathcal D_2^{(\geq 2)} +10  {\check{\mathcal D}}_3^{(\geq 2)} + 15 {\bar{\mathcal D}}_3^{(\geq 2)} + 20 {\hat{\mathcal D}}_3^{(\geq 2)}  + \sum_{k=2}^{22} \big( 5\bar{\mathcal S}_3^{(k)} + 10 \hat{\mathcal S}_3^{(k)} \big)  
\leq 66|X_2| + 115|X_3| 
\eeq
which, once added to~\eqref{336-improved} of Lemma~\ref{lem8}, results in   
\beq
r_{\rm bound} \leq 336 + 66|X_2| + 115|X_3| \,. 
\eeq
It thus follows that 
\beq
r \leq r_{\rm bound} \leq 336 + 66|X_2| + 115|X_3| \leq 336 + 115(|X_2| + |X_3|) \leq 566 \,, 
\eeq
where~\eqref{X2X3} has been used in the last step. \hfill\(\Box\)

%%%%%%%%%%%%%%%%%%%%%%%%%%%%%%%%%%%%%%%

\section*{Acknowledgements}
%%%%%%%%%%
%\section*{ACKNOWLEDGEMENTS} 
The work of S.-J.L. is supported by the Yonsei University Research Fund 2025-22-0134. C.B. is supported by a grant from Tsinghua University and a grant from National Program of Overseas High Level Talent. The authors would like to thank Fulin Xu for his helpful comments on an earlier draft of this paper.

\end{document}